\documentclass[aps,prl,floatfix,twocolumn,nofootinbib]{revtex4}
\usepackage{subfigure}
\usepackage{epic}\usepackage{eepic}
\usepackage{bm}
\usepackage{graphicx} \usepackage{amsmath} \usepackage{amssymb}
\newcommand{\comment}[1]{}
\newcommand{\BEA}{\begin{eqnarray}}
\newcommand{\EEA}{\end{eqnarray}}

\newcommand{\ta}{T_1}
\newcommand{\tb}{T_2}

\newcommand{\NA}{N_1}
\newcommand{\NB}{N_2}
\newcommand{\VA}{V_1}
\newcommand{\VB}{V_2}

\newcommand{\UA}{U_1}
\newcommand{\UB}{U_2}

\newcommand{\SA}{S_1}
\newcommand{\SB}{S_2}

\begin{document} 

\title{Thermodynamic definition of mean temperature}

\author{ A. E. Allahverdyan$^{1,2)}$, S. G. Gevorkian$^{1,3)}$, Yu. A. Dyakov$^{4)}$, and
Pao-Kuan Wang$^{4,5,6)}$ }

\address{ 
$^{1)}$Alikhanian National Laboratory (Yerevan Physics Institute), Alikhanian Brothers Street 2,  Yerevan 0036, Armenia\\
$^{2)}$Yerevan State University, 1 A. Manoogian street, Yerevan 0025, Armenia,\\
$^{3)}$Institute of Physics, Academia Sinica, Nankang, Taipei, 11529, Taiwan\\
$^{4)}$Research Center for Environmental Changes, Academia Sinica, Taipei, Taiwan,\\
$^{5)}$Department of Aeronautics and Astronautics, National Cheng Kung University, Tainan, Taiwan\\
$^{6)}$Department of Atmospheric and Oceanic Sciences, University of Wisconsin-Madison, Madison, Wisconsin, USA
}
 
\begin{abstract} The notion of mean temperature is crucial for a number
of fields including climate science, fluid dynamics and biophysics.
However, so far its correct thermodynamic foundation is lacking or
even believed to be impossible.  A physically correct definition should
not be based on mathematical notions of the means (e.g. the mean
geometric or mean arithmetic), because they ignore the peculiarities of
the notion of temperature, and because they are not unique. We offer a
thermodynamic definition of the mean temperature that is based upon the
following two assumptions. First, as the correct definition should
necessarily involve equilibration processes in the initially
non-equilibrium system, the mean temperature is bounded from below and
above via looking at (respectively) the reversible versus fully
irreversible extremes of equilibration. Second, within the
thermodynamic approach we assume that the mean temperature is determined
mostly by energy and entropy. Together with the dimensional analysis,
the two assumptions lead to a unique definition of the mean temperature.
The mean temperature for ideal and (van der Waals) non-ideal gases with
temperature-independent heat capacity is given by a general and compact
formula that (besides the initial temperatures) only depends on the
heat-capacities and concentration of gases. 
 
\end{abstract}

\maketitle 

\comment{
As a result of several important problems, it is necessary to define the
mean temperature in a non-equilibrium system. An example is the use of
the mean temperature of Earth's (surface) atmosphere as an indicator in
climate change discussions \cite{resource,collins,bjarne}. The Earth
atmosphere is certainly a non-equilibrium system, but it supposed to
admit a quasi-equilibrium description with a mean temperature. Another
example is a living body, which is a non-equilibrium system with
different temperatures for different degrees of freedom, but it
supposedly admits a single-temperature, quasi-equilibrium description,
which is employed in a macroscopic description of organisms e.g. in
physiology or medical studies; see \cite{suzuki} for recent challenges 
in this field. These two pertinent examples show that the notion of the mean
temperature is already employed in one way or another. 
}

Many non-equilibrium situations are described by mean temperature in a
quasi-equilibrium manner. Climate is defined with respect to a mean temperature,
while climate change discussions are largely based on global mean temperatures 
of Earth's surface \cite{resource,hartmann,collins,huber,west,bjarne}. Mean temperature and
deviations from it are widely used in turbulence
\cite{turbu1,turbu2,turbu3}, granular gases \cite{granular}, cellular
biophysics \cite{suzuki}, material science (including radiative heat
transfer) \cite{emig} {\it etc}. But this notion so far lacks
physical foundations \cite{bjarne}. There are several inter-related
reasons for that. 

{\it (i)} Mathematical definitions of the mean are not unique (mean arithmetic,
geometric or harmonic?), though in one way or another they
are employed for defining the mean temperature. 

{\it (ii)} The mean arithmetic is selected for additive
quantities; e.g. length, volume, and (to some extent) energy. Here
defining the mean amounts to taking the two systems with different (say)
volumes together, calculating the total volume and dividing over the
number of systems. But temperature is not an additive quantity. 

{\it (iii)} Temperature is defined with respect to a conventionally
chosen thermometer \cite{landau,chang,skow}. The readings $t_1$ and
$t_2$ of two thermometers $1$ and $2$ relates to each other via a
monotonous transformation $t_1=f(t_2)$ \cite{landau,chang,skow}. Hence
the notions of larger and smaller are well-defined for temperature, but
mathematical means are not covariant with respect to $f(x)$, in contrast
to physical quantities (energy, entropy, pressure {\it etc}) that are
invariant.  One particular example of $f(x)$ is an affine transformation
$f(x)=ax+b$, where $a$ and $b$ are constants. Three basic
scales|Celsius, Fahrenheit, and Kelvin|relate to each other via affine
transformations. The arithmetic mean (but not other means) is covariant
with respect to an affine transformations, but is not covariant with
respect to more general monotonous transformations; e.g.  to $f(x)=1/x$.
Indeed, frequently the inverse temperature $\beta=1/T$ provides a better
physical description than $T$ itself, e.g. because it provides a better
of account non-equilibrium physics \cite{landau}. Spin physics 
employs $\beta$ instead of $T$, also because in that field
$\beta$ passing through zero is usual, unlike ``dramatic'' changes
implied by $T=1/\beta$ passing through the infinity \cite{goldman}.
Also, the usage of $\beta$ (instead of $T$) makes the third law 
intuitive, since this law now tells about impossibility to reach
$\beta=\infty$ \cite{landau}. 

{\it (iv)} Temperature is defined only in equilibrium \cite{landau}.
This means that for defining a mean temperature in a non-equilibrium
state we should invoke physically meaningful equilibration processes. 
But they are not unique. 

We propose a thermodynamic definition of the mean temperature that
solves the above issues.  We start by setting upper and lower bounds to
the mean temperature.  These bounds describe two extremes of
equilibration processes: one that is reversible and thermally-isolated,
releasing work, and one that is completely irreversible and
energy-isolated, increasing entropy by dissipating the available work.
We postulate that respective temperatures $\hat T$ and $\widetilde T$
($\hat T<\widetilde T$) are lower and upper bounds of the mean
temperature $\bar T$.  Next, we assume that within the thermodynamic
description, the mean temperature $\bar T$ is defined only via entropies
and energies of the initial (non-equilibrium) state and possible final
equilibrium state.  This assumption, along with a dimensional analysis,
suffices to define the mean temperature $\bar T$.  It holds $\hat T\leq
\bar T\leq \widetilde T$, and depends both on the initial
non-equilibrium state of the considered system and also on the very
substance it refers to; e.g.  the mean temperature for two pieces of
iron having temperatures $T_1$ and $T_2$ will be different from two
pieces of wood having the same temperatures $T_1$ and $T_2$.  However,
for a class of systems relevant for atmospheric physics|ideal and van
der Waals non-ideal gases with a temperature-independent heat-capacity
(cf.~\S 1 of \cite{sm})|$\bar T$ holds a general expression that (aside
of the initial temperatures) depends only on heat-capacities and
concentrations of gases. 

\comment{
We emphasize that the proposed definition is a thermodynamic one, i.e.
it does not take into account dynamical factors, e.g. those given by
hydrodynamics or kinetic theory.  This is both strength and weakness,
also because our results become more generally applicable and
independent on dynamical details (e.g. heat-conductivity). Nonetheless,
we note that our method of placing upper and lower bounds on the mean
temperature, and then determining the latter from more fine-grained
reasoning, will be useful also in more specific, dynamic situations, 
also because those situations do assume fixed reference states \cite{atmo}. }

Consider two equilibrium systems $A_1$ and $A_2$ at different absolute
temperature $\ta$ and $\tb$, with the number of particles $\NA$ and
$\NB$, volumes $\VA$ and $\VB$, internal energies $\UA(\ta,\VA,\NA)$ and
$\UB(\tb,\VB,\NB)$, and entropies $\SA(\ta,\VA,\NA)$ and
$\SB(\tb,\VB,\NB)$, respectively.  We shall work with the absolute
temperature scale (in energy units, i.e.  with $k_{\rm B}=1$), but at
several places we emphasize the covariance of our conclusions with
respect to monotonous transformations of temperature. 

We assume that ${A_1}+{A_2}$ is a thermally isolated system. We allow
${A_1}+{A_2}$ to equilibrate and reach some joint temperature, which can
be then related to the mean temperature. The equilibration process is
not unique, it depends on the external conditions \cite{landau}; e.g.
it depends on whether and to which extent we allow for work-extraction
from ${A_1}+{A_2}$. Below we resolve this non-uniqueness.  We consider
only processes that proceed via thermal contacts; i.e. they are realized
at fixed values of the volumes $(\VA, \VB)$ and the particle numbers
$(\NA, \NB)$. The reason for this is discussed in \S 2 of \cite{sm}.
Hence we omit the arguments $(\VA,\NA)$ and $(\VB,\NB)$ for energy and
entropy. 

\begin{figure}[t]
\includegraphics[width=7cm]{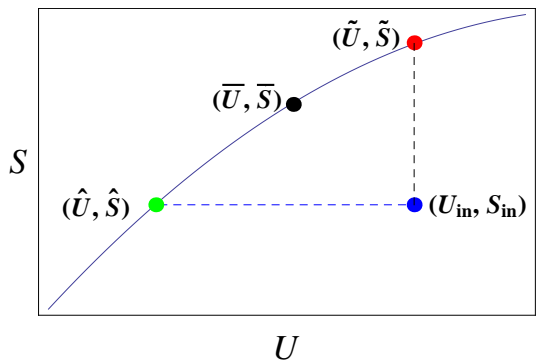}
\caption{
A schematic representation of the entropy-energy diagram. The blue
concave curve shows $S(U)$: the equilibrium entropy $S$ as a function of
equilibrium energy $U$ for a fixed volume and particle number. This
curve is growing, $S'(U)>0$, since the temperature is positive. The
curve is concave, $S''(U)<0$, because the heat-capacity (at a fixed
volume and particle number) is positive due to thermodynamic stability
\cite{landau}. \\ Now $(U_{\rm in}, S_{\rm in})$ denoted by the blue
point means the initial non-equilibrium state; e.g. this can be a
two-temperature state for two systems. This state can thermalize via (at
least) two processes: the irreversible scenario refers to a constant
energy and monotonically increasing entropy; see the black-dashed
(vertical) line. Equilibrium values are denoted as $(\widetilde U,
\widetilde S)$ (red point) and refer to temperature $\widetilde T$;
cf.~(\ref{t33}). The second process is the reversible one, where the
entropy stays constant, while the energy deceases till equilibrium
values $(\hat U, \hat S)$ (temperature $\hat T$) are reached, as in
(\ref{q33}); see the blue-dashed (horizontal) line reaching the green
point. The mean temperature $\bar T$ refers to an equilibrium state
$(\bar U, \bar S)$ located in between: $\hat T<\bar T<\widetilde T$; see
(\ref{1}). 
}
\label{fig1}
\end{figure}

{\it Reversible equilibration}. $A_1$ and $A_2$ couple
through a working body $B$, which sequentially interact with $A_1$
and $A_2$ via weak thermal contacts. Between interactions $B$ delivers
work to the external source \cite{landau}. The thermodynamic state of
$B$ changes cyclically. Hence the overall entropy change is given by the
change of the entropies of ${ A_1}$ and ${A_2}$. The process is
reversible, and the overall entropy stays constant:
\BEA
\label{qq2}
\SA (\ta)+\SB (\tb)=\SA (\hat T)+\SB (\hat T),
\EEA
where $\hat T$ is the final temperature, which is the same for $A_1$ and $A_2$. 
Since $A_1+A_2$ is thermally isolated due to (\ref{qq2}), the overall energy deficit is the extracted work: 
\BEA
\label{qq22}
\UA (\ta)+\UB (\tb)-\UA (\hat T)-\UB (\hat T)\geq 0.
\EEA
Fig.~\ref{fig1} illustrates the situation of (\ref{qq2}, \ref{qq22}) on
the energy-entropy diagram. It is seen that possible non-equilibrium
states $(U_{\rm in}, S_{\rm in})$ are bound (from left) by an an increasing and concave equilibrium
energy-entropy curve. The reason of concavity is reminded in the caption
of Fig.~\ref{fig1}. For our situation $U_{\rm in}=\UA(\ta)+\UB(\tb)$ and 
$S_{\rm in}=\SA(\ta)+\SB(\tb)$. Note from Fig.~\ref{fig1} 
that (\ref{qq22}) is the maximal work that can be extracted under the 
restriction of the second law and fixed $(\VA,\NA)$ and $(\VB,\NB)$:
\BEA
\UA (\hat T)+\UB (\hat T)={\rm min}_{ \hat{\ta}, \hat{\tb}}\left[\, \UA (\hat\ta)+\UB (\hat\tb)\,\left|.\right.\,\right],
\label{q33}
\EEA
where the minimization is conditioned by $\SA (\ta)+\SB (\tb)\leq \SA
(\hat\ta)+\SB (\hat\tb)$.  Indeed, if we allow more general processes,
where the final entropy is larger than $S_{\rm in}=\SA(\ta)+\SB(\tb)$,
then the final energy is also larger than $\UA (\hat T)+\UB (\hat T)$
due to the fact that the equilibrium energy-entropy curve $S(E)$ is
increasing; see Fig.~\ref{fig1}. 

It is natural for relaxation to be accompanied by work-extraction.
Within atmospheric thermodynamics, work-extraction means increasing the
hydrodynamic kinetic energy due to internal energy and refers to the
emergence of a macroscopic motion (wind, storm or circulation) in a
non-equilibrium state \cite{margules,lorenz}; see \cite{atmo,rmp} for
reviews.  Carefully accounting for this energy balance requires fluid
dynamic consideration; see e.g.  \cite{andrew}. Ref.~\cite{huang}
studied the maximal extracted work and the maximal entropy increase as
features of a non-equilibrium atmosphere; see \cite{rmp} for a review. 

{\it Fully irreversible equilibration}.
The second pertinent scenario of equilibration looks at another extreme.
Now $A_1$ and $A_2$ are isolated from the rest of the world and are
subject to the fully irreversible equilibration via thermal contacts; i.e.
again $(\VA,\NA)$ for $A_1$ and $(\VB,\NB)$ for $A_2$ stay fixed. Now
the total energy is conserved
\BEA
\label{tt2}
\UA (\ta)+\UB (\tb)=\UA (\widetilde T)+\UB (\widetilde T),
\EEA
defining the final temperature $\widetilde{T}$. The entropy increase is clearly positive:
\BEA
\SA (\widetilde T)+\SB (\widetilde T)-\SA (\ta)-\SB (\tb)>0.
\label{bro}
\EEA
As seen from Fig.~\ref{fig1}, (\ref{bro}) is the maximal entropy increase for the conserving energy
plus fixed $(\VA,\NA)$ and $(\VB,\NB)$:
\BEA
\SA (\widetilde T)+\SB (\widetilde T)={\rm max}_{ \widetilde{\ta}, \widetilde{\tb}}
\left[\, \SA (\widetilde\ta)+\SB (\widetilde\tb)|\,.\right],
\label{t33}
\EEA
where the maximization is conditioned by $\UA (\ta)+\UB (\tb)= \UA (\widetilde\ta)+\UB (\widetilde\tb)$.

{\it Upper and lower bounds for the mean temperature}. 
As we confirm below, the temperatures $\hat T$ and $\widetilde T$ hold
\BEA
{\rm min}[T_1,T_2]\leq \hat T\leq \widetilde{T}\leq {\rm max}[T_1,T_2],
\label{suffer}
\EEA
which naturally implies that all other temperatures found via partially irreversible processes 
will be located between $\hat T$ and $\widetilde T$; see Fig.~\ref{fig1}. 

Our first assumption reads: the mean temperature $\bar{T}$ should locate between 
$\hat T$ and $\widetilde T$:
\BEA
\label{a1}
\hat{T}\leq \bar{T}\leq \widetilde T
\EEA
The motivation for (\ref{a1}) is that once the temperature relates to
the heat content, we should decide what to do with the available work.
Two extreme options are to dissipate it completely ($\widetilde T$), or
to extract it fully ($\hat T$). We cannot add any external work, since
this will potentially change the heat content. Moreover, if we start
adding work to the overall system ${ A_1}+ { A_2}$, the final
temperature can be made arbitrary large. 

We emphasize that $\hat{T}$ and $\widetilde T$ are covariant with respect 
to monotonous transformations of the absolute temperature. This follows from 
the very definitions (\ref{qq2}, \ref{tt2}). Put differently, $\hat{T}$ and $\widetilde T$
are covariant with respect to employing any other reasonable thermometer instead of the 
thermometer that leads to the absolute temperature.

\begin{figure}[t]
\includegraphics[width=7cm]{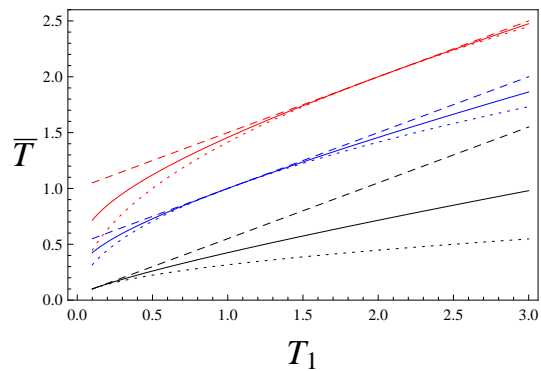}
\caption{
The mean temperature $\bar T$ for a two-temperature system {\it versus} temperature $T_1$ of one of systems; see 
(\ref{osh}). From top to bottom (red, blue, black curves): $T_2=2,\,1,\, 0.1$. For each curve its dashed (dotted) counterpart
of the same color denotes the arithmetic mean $\widetilde T=(T_1+T_2)/2$ (geometric mean $\hat T=\sqrt{T_1T_2}$) 
calculated at the same value of $T_2$. It is seen that $\widetilde T\geq \bar T\geq \hat T$.
}
\label{fig2}
\end{figure}

{\it Mean temperature}.
Guided by (\ref{q33}, \ref{t33}), we postulate that the mean temperature $\bar{T}$ is 
found through a four-variable function of the initial and final energy 
and entropy:
\BEA
\label{ur}
\bar{T}={\rm argmax}_{T}
F[\UA ( T)+\UB ( T), \SA ( T)+\SB ( T), \nonumber\\
\SA (\ta)+\SB (\tb), \UA (\ta)+\UB (\tb)]. 
\EEA
$\bar T$ in (\ref{ur})
should be invariant with respect to changing
dimensions of the entropy and energy, as well as adding to these
quantities arbitrary constants $b$ and $d$ \cite{landau}:
\BEA
\label{ura}
S\to a S+b, \qquad U\to c U+d,\qquad a>0,\quad c>0,
\EEA
where the constants $a$ and $c$ come from changing the dimensions. In addition, 
(\ref{ur}) should hold (\ref{a1}). 
Indeed, the invariance with respect to changing the dimensions is a
natural condition to demand, while the invariance with respect adding
arbitrary constants reflect the freedom energy and entropy enjoy in
thermodynamics. 

The only expression that holds all above conditions, and is invariant
with respect to (\ref{ura}) reads \cite{zang,roth,game}:
\begin{gather}
\bar{T}(\alpha_1, \alpha_2)={\rm argmax}_T\{\,
[\,\UA (\ta)+\UB (\tb)-\UA ( T)-\nonumber \\ \UB ( T)]^{\alpha_1} \,
[\,\SA ( T)+\SB ( T)-\SA (\ta)-\SB (\tb)]^{\alpha_2} \},
\label{10}
\end{gather}
where $\alpha_1\geq 0$ and $\alpha_2\geq 0$ are weights. All other possibilities are equivalent to
(\ref{10}) in one way or another. 

Note that for $\alpha_1\to 0$ and $\alpha_2\to 0$ we revert to (resp.) 
(\ref{t33}) and (\ref{q33}). It should be also clear that $\hat T\leq \bar{T}(\alpha_1, \alpha_2)\leq \widetilde T$.
How to choose $\alpha_1$ and $\alpha_2$? First note that
instead of the function to be maximized in (\ref{10}) we can maximize its
logarithm, which makes clear that only the ratio
$\alpha_1/\alpha_2$ is important for $\bar{T}(\alpha_1, \alpha_2)$. In
other words, we can assume $\alpha_1+\alpha_2=1$. Once we do not
know these weights, the ignorance (or the most unbiased, or the maximum
entropy) interpretation forces us to choose $\alpha_1=\alpha_2=1/2$.
Thus we end up from (\ref{10}) to the final definition of the mean
temperature:
\BEA
\bar{T}=\bar{T}(1/2, 1/2).
\label{1}
\EEA
Eq.~(\ref{1}) achieves a balance between no work extraction (complete
irreversibility) and the complete reversibility: now some work is still
extracted, but the entropy does increase. The unique argmax of (\ref{1}) 
automatically appears in the interval $[\hat T, \widetilde T]$. 
In contrast to a naive definition $(\widetilde T+\hat T)/2$, $\bar T$ in
(\ref{1}) is invariant with respect to monotonic changes of temperature
(i.e. going from one reasonable thermometer to another), since it is
defined via optimization of thermodynamic variables. 

To extend the physical meaning of (\ref{1}, \ref{10}) note that 
$\bar T$ from (\ref{1}) is the temperature that corresponds to values
$(\bar U, \bar S)$ found via
\begin{gather}
(\bar U, \bar S)={\rm argmax}_{S,U}\left[(S-S_{\rm in})(U_{\rm in}-U)  \right], 
\label{111}
\end{gather}
where the maximization is carried out over all allowed physical (also
non-equilibrium) values of $S$ and $U$. For consistency with (\ref{1}) we have $S_{\rm
in}=\SA(\ta)+\SB(\tb)$, and $U_{\rm in}=\UA(\ta)+\UB(\tb)$, but
(\ref{111}) applies to any non-equilibrium initial state $(S_{\rm in},
U_{\rm in})$; e.g. several initially non-interacting, equilibrium
systems at different temperatures, or non-equilibrium systems with 
effective temperatures \cite{glass,onsager,jou,epl,sobolev}; 
cf.~\cite{marti2}. To understand (\ref{111}), recall
from Fig.~\ref{fig1} that those physical values of energy and entropy
are bound into a convex domain by the equilibrium curve $S(U)$. The
maximization in (\ref{111}) is reached in that curve; otherwise one can
always increase $S$ or decrease $U$ so as to reach this curve. We also
naturally have $\hat S\leq \bar S\leq \widetilde S$ and $\hat U\leq \bar
U\leq \widetilde U$; see (\ref{111}) and Fig.~\ref{fig1}.  The maximum
in (\ref{111}) [i.e. also in (\ref{1})] is unique; see \S 3 of
\cite{sm}.  Eq.~(\ref{111}) can be applied without demanding that $(\bar
U, \bar S)$ are in equilibrium, and without requiring the bounds
(\ref{a1}). These features come out automatically from (\ref{111}),
which can be generalized to cases where there are additional 
(dynamic) restrictions towards attaining the equilibrium curve.

{\it Ideal gases}.  The simplest case is that of
$M$ ideas gases with volumes $V_k$, number of particles $N_k$, internal
energies $U_k$, entropies $S_k$, {\it constant} (i.e.
temperature-independent) fixed-volume heat-capacities $c_k$, and
temperatures $T_k$ \cite{landau} ($k=1,..,M$):
\BEA
\label{o1}
S_k=N_k\ln\frac{V_k}{N_k}+N_kc_k\ln T_k,\qquad U_k=N_kc_kT_k,
\EEA
where in $S_k$ and in $U_k$ we omitted certain inessential constants. Now $\hat T$ and $\widetilde T$
read from (\ref{q33}, \ref{t33}):
\begin{gather}
\label{ost1}
\hat T={\prod}_{l=1}^M T_l^{n_lc_l/\bar{c}},\quad 
\widetilde T={\sum}_{l=1}^M \frac{n_lc_l}{\bar{c}} T_l,\\
\bar{c}={\sum}_{l=1}^Mn_lc_l,\quad n_l=\frac{N_l}{N}, \quad N={\sum}_{l=1}^M {N_l},
\end{gather}
where $\bar{c}$ is the mean heat-capacity and $n_l$ are concentrations.
Hence $\hat T$ ($\widetilde T$) in (\ref{ost1}) reduces to a weighted
geometric (arithmetic) average with weights $n_kc_k/\bar{c}$. The
arithmetic and geometric means in (\ref{ost1}) are found only for ideal
gases having temperature-independent heat-capacities, i.e. they do not
hold generally. 

Now (\ref{1}) is represented as ${\rm max}_T\left[(\widetilde T-T) \ln({T}/{\hat T})\right]$. The maximization is 
carried out by differentiating:
\BEA
\label{faso}
\bar T=\widetilde T\left/ W\left[e\,{\widetilde T}/{\hat T} \right]  \right.,
\EEA
where $W[z]$ is the Lambert special function that solves the equation $We^W=z$ \cite{canadian}. 
It has various applications in physics \cite{canadian}, and is tabulated with major computational 
platforms; e.g. Python and Mathematica (as ${\rm PolyLog}[z]$). 
In the simplest case of equal heat-capacities, $c_1=...=c_M$, we find from (\ref{faso}) for $M=2$ (with obvious 
generalization to $M>2$):
\BEA
\bar T=\frac{T_1+T_2}{2 W\left[e\,(T_1+T_2)/(2\sqrt{T_1T_2})\right]}.
\label{osh}
\EEA
Fig.~\ref{fig2} shows the behavior of (\ref{osh}) along with $\hat T$ and $\widetilde T$.

The same formulas (\ref{ost1}--\ref{osh}) apply also to the case, where
each system $k$ ($k=1,...M$) at temperature $T_k$ is not a single ideal
gas but a mixture of ideal gases.  Then the only change in
(\ref{ost1}--\ref{osh}) is that $c_k$ ($k=1,...M$) is the average
(temperature-independent) heat-capacity of the mixture.  This case
refers to the mean temperature of two stations that measured
temperatures $T_1$ and $T_2$ (resp.) for air with different composition
of the main atmospheric gases: nitrogen, oxygen and argon. This
difference translates into different values of $c_1$ and $c_2$, and then
(\ref{ost1}--\ref{osh}) apply. Importantly, the same
(\ref{ost1}--\ref{osh}) apply for van der Waals non-ideal gases; see \S
1 of \cite{sm}. 

Eqs.~(\ref{ost1}--\ref{osh}) do not apply if the heat-capacity $c$
depends on temperature. Here we should proceed directly from
(\ref{1}). This temperature-dependence is small for air: $c$ changes by $0.4\%$ for
temperature $T$ varying between 300 K and 350 K. Thus (\ref{ost1}--\ref{osh})
directly apply to calculating the mean temperature of air. 

{\it In sum,} multiple problems arise when mean temperature is used to
explain inhomogeneous temperature situations; see {\it (i-iv)} in the
introduction. These problems will likely become more serious, once
temperature methods go deeper into micro-scales
\cite{karen,marti,marti2}. We proposed a solution to these problems
based on thermodynamics. It is sufficiently flexible to include
additional microscopic constraints coming e.g. from fluid dynamics.

\begin{acknowledgments} We thank Hakob Avetisyan whose remark initiated this work. We
acknowledge discussions with Vardan Bardakhchyan. We thank B. Andresen for critical remarks. 
This work was supported by SCS of Armenia, grants No. 20TTAT-QTa003 and No. 21AG-1C038. 
\end{acknowledgments}

\clearpage

\section*{\large Supplementary Material for\\ {\it {Thermodynamical
definition of mean temperature}}\\ by A.E. Allahverdyan et al.}

This Supplementary Material consists of 3 chapters. \S 1 reminds
thermodynamics of van der Waals non-ideal gas and shows that the
formulas for the mean temperature hold also for this gas. \S 2 explain
why we restricted thermalization processes in the main text. \S 3
discusses a technical issue related to the uniqueness in optimization.
Formulas continue the numbering of the main text. 

\section*{\S 1. van der Waals non-ideal gas }

For the van der Waals non-ideal gas model entropy and energy read, respectively
\cite{landau1} [cf.~(\ref{o1})]:
\BEA
\label{o5}
S_k=c_kN_k\ln T_k+N_k\ln\left[ \frac{V_k-N_kb_k}{N_k} \right],\\
U_k=c_kN_kT_k-\frac{N_k^2a_k}{V_k},\qquad k=1,...,M,
\label{o555}
\EEA
where $c_k$ is the constant-volume heat capacity|which are
again assumed to be constants|while 
$a_k$ and $b_k$ are van der Waals parameters. Recall that $a_k>0$ enters
only into the energy, i.e. it characterizes the inter-particle 
interaction, while $b_k>0$ enters only the entropy, as it stands for
the excluded volume. It should be clear from (\ref{o5}) that 
$\frac{N_kb_k}{V_k}$ holds $0<1-\frac{N_kb_k}{V_k}$. An upper bound  
on $1-\frac{N_kb_k}{V_k}$ comes from the thermodynamic 
stability condition $\frac{\partial P}{\partial V}|_T<0$
\cite{landau1}. Hence two bounds together can be written as:
\BEA
0<1-\frac{N_kb_k}{V_k}<\sqrt{\frac{T_k}{2a_k\frac{N_k}{V_k}}}.
\EEA

Given these restrictions on the van der Waals parameters we find back from (\ref{1}, \ref{10}) the same formulas
(\ref{ost1}--\ref{osh}). In particular, they apply for defining averages between metastable
states described by the van der Waals equation. 

\section*{\S 2. Why we do not consider more general equilibration processes?}

In the main text we restricted ourselves with reversible and fully
irreversible processes that proceed via thermal contacts; i.e. they are
realized at fixed values of the volumes $(\VA, \VB)$ and the particle
numbers $(\NA, \NB)$. In particular, we did not involve pressure
differences into the work-extraction, because even when the equilibrium
systems $A_1$ and $A_2$ have initially the same temperature $T$, their
final temperature (after work has been extracted from pressure
differences as well) will be lower than $T$, as we show below.  This
would obviously contradict our intention of defining the mean
temperature, since e.g. condition (\ref{suffer}) will not anymore hold.
Similar issues arise when $A_1$ and $A_2$ are composed of different
(distinguishable) particles, and we allow the mixing of gases during the
work-extraction. Then the final temperature will be lower than $T$ due
to the Gibbs mixing term even when initial pressures and temperatures
are equal; see \S 2.1 below. 

This issue is not restricted to the reversible mode of operation only.
For example, during an irreversible mixing of two {\it non-ideal} gases
$A_1$ and $A_2$ having initially the same temperature $T$, their final
temperature will be lower than $T$, if the irreversible process is
extended to include pressure differences; see \S 2.1. 

Thus, the definition of a mean temperature in a non-equilibrium system
requires equilibration processes that proceed mostly thorough thermal
conductivity. \S 2.1 and \S 2.2 show that there are reversible and
irreversible processes that are not restricted to thermal contacts and
that are not suitable for determining the mean temperature for a number
of interesting reasons. 

\subsection*{\S 2.1 Ideal gases}

Recall (\ref{o1}) for entropy and energy of $k=2$ ideal gases at (initial) temperatures $T_k$. 
The two gases together form a thermally isolated system. The total 
volume $V_1+V_2$ and the total number of particles $N_1+N_2$ are conserved. 

First we assume that $c_1\not=c_2$, i.e. the gases are different. Let
now they mix together and equilibrate via an entropy conserving process.
Hence only $V_1+V_2$ and $N_1+N_2$ are conserved, but not $V_1$ and
$N_1$ separately. In the final equilibrated state each gas
occupies volume $V=V_1+V_2$ and they both have the same temperature
$\hat T$. Hence the condition that the final entropy equals initial
entropy reads from (\ref{o1})
\begin{gather}
\sum_{k=1}^2N_k\left(\ln\frac{V}{N_k}+c_k\ln \hat T\right)=
\sum_{k=1}^2N_k\left(\ln\frac{V_k}{N_k}+c_k\ln  T_k\right).
\label{x2}
\end{gather}
Now (\ref{x2}) leads to
\begin{gather}
\label{x5}
\hat T=e^{-\frac{1}{\bar{c}} \sum_{l=1}^2n_l\ln\frac{1}{v_l} }\prod_{l=1}^2 T_l^{n_lc_l/\bar{c}},\\
n_k=\frac{N_k}{\sum_{l=1}^2 N_l }, \qquad v_k=\frac{V_k}{\sum_{l=1}^2 V_l }, \qquad 
\bar{c}=\sum_{l=1}^2 n_lc_l,
\label{x55}
\end{gather}
where $\sum_{l=1}^2n_l\ln\frac{1}{v_l}\geq 0$ is the Gibbs mixing term. We now have
for $T_1=T_2$:
\BEA
\hat T=e^{-\frac{1}{\bar{c}} \sum_{l=1}^2n_l\ln\frac{1}{v_l} }<T_1=T_2, 
\EEA
i.e. this equilibration 
scheme is not suitable for the definition of a mean temperature. 

Now assume that the gases are identical, hence $c_1=c_2=c$. The final
equilibrated gas should be treated as a single entity with number of
particles $N=N_1+N_2$, temperature $\hat T_{\rm id}$ and volume
$V=V_1+V_2$. The constant entropy condition reads instead of (\ref{x5})
\BEA
\label{x6}
N\ln\frac{V}{N}+c\ln \hat T_{\rm id}=
\sum_{k=1}^2N_k\left(\ln\frac{V_k}{N_k}+c\ln  T_k\right),
\EEA
which implies [cf.~(\ref{x55})]
\BEA
\hat T_{\rm id}=e^{-\frac{1}{{c}} \sum_{l=1}^2n_l\ln\frac{n_l}{v_l} }\prod_{l=1}^2 T_l^{n_l}.
\label{x7}
\EEA
Recall from the fact that $\sum_{l=1}^2n_l\ln\frac{n_l}{v_l}$ is a relative entropy:
\BEA
\label{relo}
\sum_{l=1}^2n_l\ln\frac{1}{v_l}\geq \sum_{l=1}^2n_l\ln\frac{n_l}{v_l}\geq 0, 
\EEA
and $\sum_{l=1}^2n_l\ln\frac{n_l}{v_l}= 0$ only when $n_l=v_l$
($l=1,2$).  It is seen again that even when $T_1=T_2$ in (\ref{x7}) we
still have that $\hat T_{\rm id}<T_1=T_2$. Noting the equation of state
$P_kV_k=N_kT_k$ (where $P_k$ is pressure and $k=1,2$) for initial gases,
we see that for $T_1=T_2$ we can have
$\sum_{l=1}^2n_l\ln\frac{n_l}{v_l}> 0$ only when $P_1\not=P_2$. Hence
the inequality of pressures in the initial state makes $\hat T_{\rm id}<T_1=T_2$. 

Thus, due to $\hat T_{\rm id}<T_1=T_2$ and $\hat T<T_1=T_2$, processes described 
by (\ref{x5}, \ref{x6}) are not suitable for defining a lower bound
on the mean temperature. 

\subsection*{\S 2.2 Non-ideal gases}

The analysis of \S 2.1 will be repeated for non-ideal gases, since there
novel points in this case. 

Recall from (\ref{o5}) the van der Waals non-ideal gas model. For
simplicity we take $c_k=c>0$ (temperature-independent constant),
$a_k=a>0$, $b_k=b>0$, and $k=1,2$. Let us assume that the two gases mix
in a completely irreversible way and reach temperature $\widetilde T$
(or $\widetilde{T}_{\rm id}$) that is determined from the energy
conservation:
\BEA
\label{j1}
\widetilde{T}=\sum_{k=1}^2 n_kT_k-\frac{aN}{cV}\sum_{k=1}^2 n_k(\frac{n_k}{v_k}-1), \\
\widetilde{T}_{\rm id}=\sum_{k=1}^2 n_kT_k-\frac{aN}{cV}\sum_{k=1}^2 n^2_k(\frac{1}{v_k}-1),
\label{j2}
\EEA
where $n_k$ and $v_k$ are defined by (\ref{x55}). Eq.~(\ref{j1}) refers
to the distinguishable situation (non-identical gases), where in the
final state we still have 2 gases with particle numbers $N_1$ and $N_2$
occupying volume $V=V_1+V_2$. Eq.~(\ref{j2}) refers to identical gases,
where the final state is a single gas at volume $V=V_1+V_2$, paricle
number $N=N_1+N_2$ and temperature $\widetilde{T}_{\rm id}$. We
emphasize that for non-ideal gases also the irreversible mixing
temperature starts to feel whether the particles are identical. 

Note from (\ref{j1}, \ref{j2}) that
\BEA
\label{j3}
\widetilde{T}<\sum_{k=1}^2 n_kT_k, \qquad \widetilde{T}_{\rm id}<\sum_{k=1}^2 n_kT_k.
\EEA
The second relation in (\ref{j3}) is obvious from $v_k<1$. The first relation follows 
from (\ref{relo}) upon noting there $\ln\frac{n_l}{v_l}\leq \frac{n_l}{v_l}-1$. Eq.~(\ref{j3})
confirms that $T_1=T_2> \widetilde{T}$ and $T_1=T_2> \widetilde{T}_{\rm id}$, i.e.
this irreversible process is not suitable for defining an upper bound on the mean temperature.


\section*{\S 3. The maximum in (\ref{111}) is unique.}

Let us show that the maximum in (\ref{111}) [hence also in (\ref{1})] is unique.
The mathematical structure of this argument is taken from \cite{roth1}; cf.~also \cite{game1}.
Indeed, if the maximum is reached at two different points $( U_1, S_1)$ and $( U_2, S_2)$ such that 
\BEA
(S_1-S_{\rm in})(U_{\rm in}-U_1)=(S_2-S_{\rm in})(U_{\rm in}-U_2), 
\EEA
then $\frac{1}{2}(U_1+ U_2, S_1+S_2)$ is also in the maximization domain
of (\ref{111}), because that domain is a convex set. This fact follows
from the concavity of $S(U)$ curve in Fig.~\ref{fig1}. Now
\begin{gather}
\frac{1}{4}(S_1-S_{\rm in}+S_2-S_{\rm in})(U_{\rm in}-U_1+U_{\rm in}-U_2)>\nonumber\\
(S_1-S_{\rm in})(U_{\rm in}-U_1)
=(S_2-S_{\rm in})(U_{\rm in}-U_2),
\end{gather}
contradicts to the assumption that $(S_1-S_{\rm in})(U_{\rm in}-U_1)=(S_2-S_{\rm in})(U_{\rm in}-U_2)$
provide maxima. Hence, the maximum in (\ref{111}) is unique: $( U_1, S_1)=( U_2, S_2)$.

\end{document}